\documentstyle[festkoer,psfig]{vieweg}

\author{J. K\"{o}tzler}
\Kurzautor{J. K\"{o}tzler}
\title{Nucleation of Stable Superconductivity in YBCO-Films}
\Kurztitel{Nucleation of Stable Superconductivity in YBCO-Films}
\Adresse{Angewandte Physik und  Zentrum
f\"{u}r Mikrostrukturforschung, Universit\"{a}t Hamburg}
\begin{document}
\bibliographystyle{prsty}
\Titel

\begin{abstract} 
By means of the linear dynamic conductivity, inductively measured on epitaxial films between 30mHz and 30 MHz, 
the transition line $T_g (B)$ to generic superconductivity is studied in fields between B=0 and 19T. It follows closely the melting line $T_m (B)$ described recently in terms of a blowout of thermal vortex loops in clean materials. The critical
exponents of the correlation length and time near $T_g (B)$, however, seem to be dominated by some intrinsic disorder. 
Columnar defects produced by heavy-ion irradiation up to field-equivalent-doses of $B_{\phi} =  10T$ 
lead to a disappointing reduction of $T_g (B \to 0)$ while for $B>B_{\phi}$  the generic line of the pristine film
is recovered. These novel results are also discussed in terms of a loop-driven destruction of generic superconductivity.
\end{abstract}

\section{Introduction}
Due to the complex electronic and real lattice structure of the high-$T_c$ cuprates
the nucleation of long-range superconductivity is still heavily debated. In the first place, 
this applies to the pairing mechanism in a non-Fermi liquid in a doped Mott--insulator \cite{A97} 
giving rise to the formation of a local superfluid density near some mean field temperature $T_0$. 
Second, the appearance  of generic superconductivity, i.e. of a superfluid density $n_s$(T) and a
vanishing linear resistance requires a rigid coupling of the phases of the local droplets. 
For high-$T_c$ materials, the effect of the low carrier concentration on the generic transition 
line has been emphasized recently \cite{EK95}. As illustrated by Fig.1a, an estimate of the phase 
fluctuations within the 3D XY-model leads to an upper bound for superconductivity

\begin{equation}
k_B T_{\Theta} = \epsilon _0 d_c  .
\label{Eq1}
\end{equation}

The rigidity against phase fluctuations is determined by a microscopic cutoff length $d _c$ \cite{EK95}
for the fluctuations perpendicular to the $CuO_2$-planes and by the stiffness parameter
$\epsilon_0 = \hbar^2 /2m^{\ast}\cdot n_s (T) = \Phi_0 ^2/(4 \pi \mu_0 \lambda^2)$ ,
which by using the flux quantum, $\Phi_0 = h/2e$,  can be determined from 
the penetration depth $\lambda$ . In this picture, between $T_0$ and $T_c$ a broad paracoherent 
regime arises for underdoped materials, $\delta  < \delta _{opt}$ , which is related to the widely 
evidenced pseudogap phenomena.
\begin{figure}[tb]
\parbox{80mm}{\psfig{figure=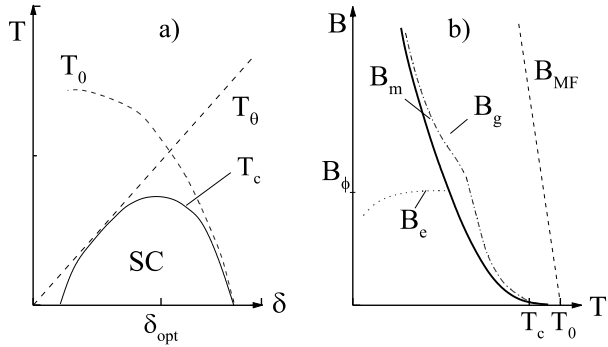,width=80mm}}  \parbox{40mm}{
\begin{minipage}[t]{40mm}{\caption{Sketches for phase diagrams of high-$T_c$ superconductors:
transition temperatures as functions of a) hole doping {\protect{\cite{EK95}}} in zero magnetic field
and b) in magnetic field near optimal doping. Dashed and dotted lines are described in the text.}}
\end{minipage}}
\label{FIG1}
\end{figure}
In real situations, several constraints may enhance the phase fluctuations 
and thus further reduce $T_{\Theta} $. At the high temperatures of interest, thermal fluctuations decrease
the stiffness $n_s$ while finite magnetic fields - the earth´s field is sufficient for thin
films due to their strong demagnetization - induce a vortex system, which 
has to be pinned in order to avoid lossy flux creep even under applied low current density. 
The issue of nucleation of generic superconductivity in the presence of current densities
and of vortices pinned by various kinds of disorder has been reviewed from the theoretical side~ \cite{FFH91, BFGLV94,VKZ98}. Today there is ample experimental evidence \cite{SFP96} for the thermodynamic melting line of the Abrikosov vortex lattice $B_m (T)$ in clean materials with different anisotropy, $\gamma = \xi_{ab} / \xi_c$
(Fig.1b). The location of $B_m (T)$ is described in terms of the Lindemann
criterium 

\begin{equation}
kT = \epsilon _0 (T) c_L^2 \frac{a_0}{\gamma} 
\label{Eq3} 
\end{equation}
where $a_0 = \sqrt { \Phi_0 / B} $ is the mean vortex spacing for $B\perp$ to $CuO_2$ planes. The comparison with Eq.(\ref{Eq1})  indicates that the melting 
is determined by the elastic cutoff $c_L^2 a_0 / \gamma $ given by the small  Lindemann number 
$c_L$.  Recent work \cite{VKZ98} reveals that random point defects give
rise to a quasi-lattice (Bragg-glass) at low fields $(B<B_e)$. Above $B_e$  entangled vortices (vortexglass, VG) are
expected due to the increasing repulsive vortex interaction, which drives the vortices into the
local energy minima set by the fluctuations of the point defects, however, generic superconductivity, as
conjectured previously \cite{FFH91} for the VG, could not yet be distinguished from vortex fluctuations just 
frozen by uncorrelated disorder \cite{OSZ98}.

For correlated disorder, on the other hand, like columnar defects \cite{C97}piercing the entire sample along $\vec{B} $, 
the existence of a generic phase transition $B_g (T)$ (Fig.1b) has been evidenced by means of the boson
mapping of the vortex lines\cite{NV93}. Accordingly, the socalled Bose-glass (BG) phase arises
from a localization of interacting 2D-bosons in a random potential. It has been argued, that not only
defects introduced by heavy ion irradiation but also intrinsic linear structures along $\vec{B} $ arising e.g. 
by edges,  twin colonies or grains of different orientation, or screw dislocations may lead to BG-ordering \cite{NV93}.

As for the melting line $B_m$, numerous evidence for generic superconductivity mostly for YBCO has been obtained using various 
experimental techniques. However, a detailed understanding of the location of 
$B_g (T)$ in terms of the pinning mechanism in a concrete material, even for well-defined columnar defects,
has not yet been achieved. In this work, we investigate $B_g (T)$ in c-axis oriented, granular YBCO films with large
critical current density, ${j_c(T \rightarrow 0) = 2\cdot {10^7}A/cm^2}$ grown for SQUID fabrication \cite{SGM94}. In addition to the intrinsic disorder, we study
the influence of columnar defects of various density in order to compare intrinsic with 
correlated pinning. We also investigate the singularities of the fluctuation conductivity
near $T_g (B)$ in order to shed more light into the nucleation process of generic superconductivity in real high-$T_c$
materials.  

\section{Inductive Probe of Generic Superconductivity}

Along with direct, four terminal I/V - and ac-probes of the superconducting transition, for the brittle cuprates
inductive techniques became very popular, because they avoid contacts and turn out to be high
sensitive to the high conductances near $T_c$ of films or bulk samples. Most of them employ
irreversibility effects in the magnetization, the screening, the onsets of nonlinearity or of higher order 
harmonics in the ac-susceptibility $\chi'$, and maxima in the absorption $\chi''$ to detect $T_g (B)$.
Similar as with the voltage criterion in I/V-curves, commonly used to define the transition temperature,
also these methods are based on at least one extrinsic condition, like frequency, measuring time, 
excitation amplitude, sample size, or simply the detection limit.

Here we use the frequency dependence of the magnetic susceptibility measured between
30mHz and 30MHz with amplitudes $b_0 (\omega )$ from $10^{-4}$ to $10^{-1} mT$ to keep 
within linear response and to optimize the sensitivity. Fig.2a illustrates that
the current density and hence the conductivity $\sigma  (\omega, T )$ in the $CuO_2$ planes is
probed. For long cylinders, where demagnetizations are negligible, $\chi (\omega )$ can directly be 
obtained by integrating the solutions of Helmholtz' diffusion equation, $\dot{\vec{b}}  = (\mu _0 \sigma )^{-1} \Delta \vec{b}$
as used e.g. in Ref. \cite{KKNBA94}.  The self-field effects produced by thin films, lead to  
a non-local diffusion equation, which has been solved numerically for all
practical sample geometries and field configurations by E.H. Brandt  \cite{B94}. The results are presented in the form

\begin{equation}
\chi (\omega )=k^2 (\omega ) \sum_n  \frac {c_n / \Lambda_n }{k^2(\omega)+ \Lambda_n (1+2 \pi R / L_z)}  
\label{Eq4}  
\end{equation}
where the coefficients $c_n$ and $\Lambda _n$ become independent on sample size for $L_z << R$. 
The complex conductivity in $k^2 (\omega ) = i\omega R^2  \mu_0 (\sigma' (\omega )  + i \sigma'' (\omega ) ) $ is extracted from $\chi (\omega )$ 
using a inversion routine. For the present precision, $\delta \chi / \chi \approx  10^{-4}$, n = 30 terms are
sufficient, which for our case of circular discs have been tabulated in Ref. \cite{KNBBGB94}, where
also more details are given.
\begin{figure}[t]
\begin{center}
\psfig{figure=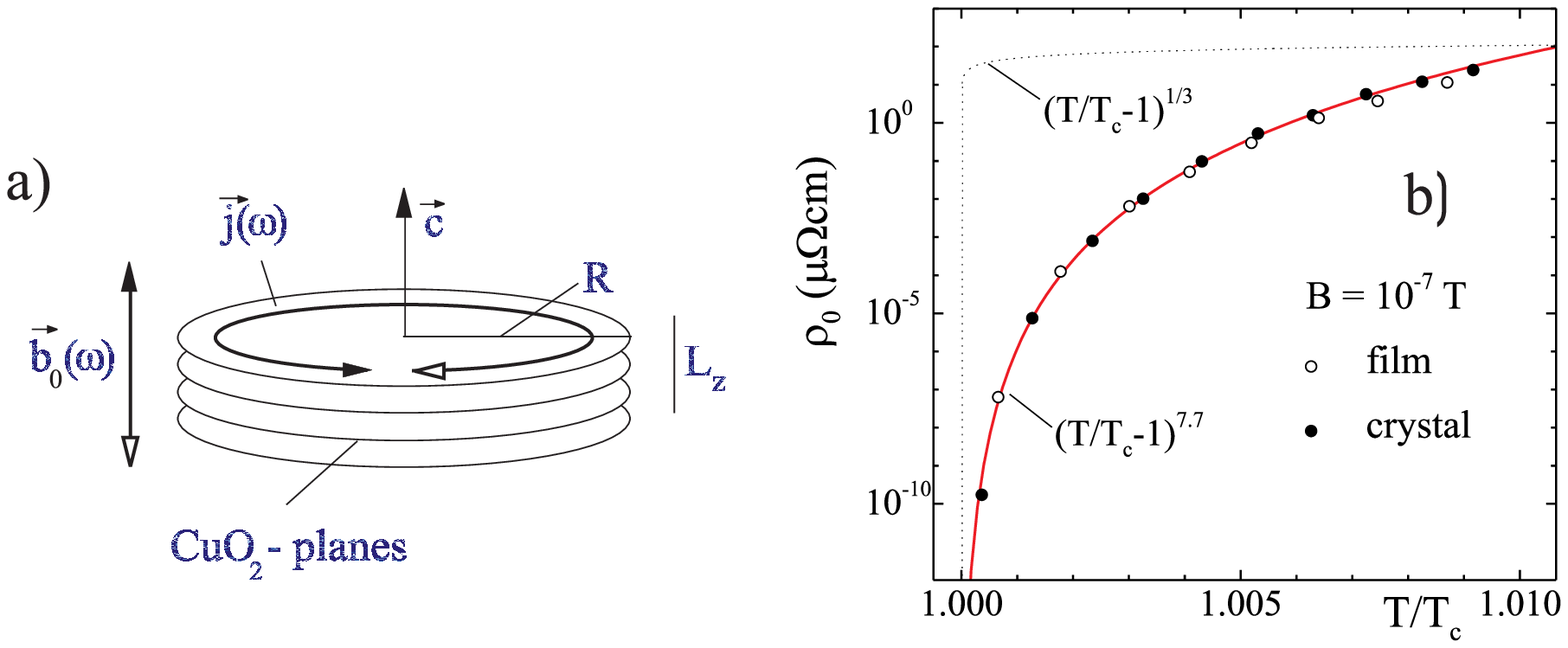,width=110mm}
\hfill
\caption{a) Contactless technique for measuring high resistivity 
parallel to the $CuO_2$ planes;  b) dc-resistivity measured in zero field near 
$T_c$ on a heavily twinned $YBa_2Cu_3O_{7-\delta} $ crystal  ($L_z =800\mu m$) and a thin film ($L_z =0.2\mu m$).
Indicated are power law singularities, $ \rho \sim (T/T_c  -1)^s$ , with s=7.7 fitting the data and $s=1/3$ (dotted) predicted 
for clean superconductors. }
\label{FIG2}
\end{center}
\end{figure}

As example, we compare in Fig.2b the zero-field dc-resistivities of the present thin film, $L_z = 0.2 \mu m$, 
with that of a heavily twinned crystal, $L_z = 800 \mu m $, \cite{KKNBA94}. Despite the different $T_c 's$, 
88K and 91K, the results fall on the same power law, which is characteristic for a
continuous phase transition. This also indicates some universality, which, however, deviates significantly from
the prediction of models for {\em clean}  materials: the mean-field, s = 1/2, and the 3D-XY with s = 1/3.
\begin{figure}[t]
\parbox{70mm}{\psfig{figure=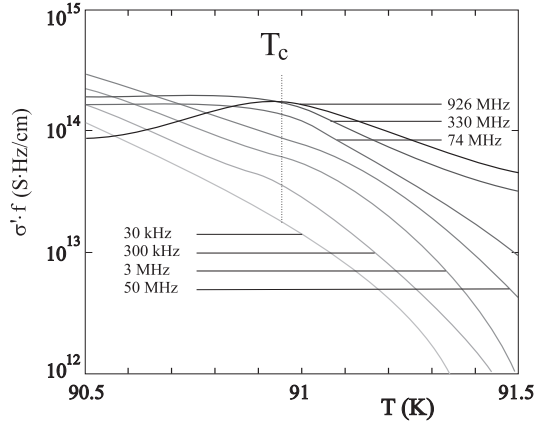,width=70mm}}  \parbox{50mm}{
\begin{minipage}[t]{50mm}
{\caption{ Real part of the dynamic conductivity multiplied by frequency  near the transition $T_c = 90.95 K$ (s. Fig.4) to generic (stable) superconductivity of a $YBa_2Cu_3O_{7-\delta}$  
film in zero magnetic field.}}
\end{minipage}}
\label{FIG3}
\end{figure}

\begin{figure}[b]
\begin{center}
\psfig{figure=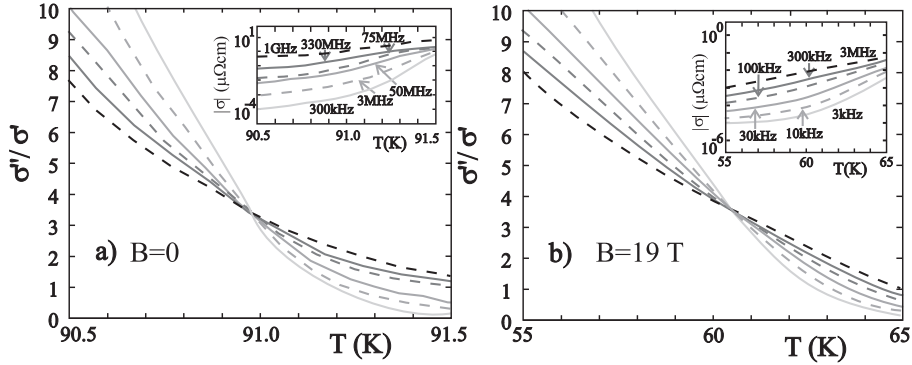,width=120mm}
\hfill
\caption{Temperature dependence of the phase angle of the dynamic conductivity,  $\sigma ' - i \sigma ''$ a) in zero magnetic field 
determined from the data of Fig.3 and b) at $B=19T$  for the same film. Insets display the modulus $|\sigma|$.}
\label{FIG4}
\end{center}
\end{figure}

The ac-conductivity near $T_c$, as shown for $\sigma'$ in Fig.3, reveals no clear signature
of a sharp  transition.  The values 
of $\sigma$ are rather high as compared to the Drude background~\cite{WNKSJHRK98}. They arise from the low frequencies
which indicates strong superconducting fluctuations,
\begin{equation}
\sigma (\omega ,T) = \frac{1}{\mu_0 \lambda_p^2} \left\{ \frac{n_s(T)}{ i \omega} + \tau_n \right \} +\sigma_{fl}(\omega,T).
\label{Eq5}
\end{equation}
In the first, the 2-fluid term, $\lambda _p ^2 = m^{\ast}/\mu_0 e^2 n_e$ denotes the plasma wavelength and 
$\tau _n^{-1}= 2kT_c/\hbar  = 2.5 \cdot  10^{13} s^{-1}$ \cite{WNKSJHRK98}
the large scattering rate of the quasiparticles. In the absence of fluctuations, the phase angle $\sigma ''/\sigma '$
provides the strongest signal at $T_c$ by jumping from zero to $n_s/\omega \tau _n$. Due to $\sigma_{fl}$ this jump
is smeared by screening and lossy contributions appearing above and below $T_c$, respectively. They produce a  positive Drude-like
frequency gradient $d(\sigma ''/\sigma ')/d\omega >0$ above $T_c$ and the inverse behaviour, $d(\sigma ''/\sigma ')/d\omega <0$ below $T_c$, so that for not too high frequencies, $\omega \tau _n <<1$,
at $T_c$ a crossing of all $\sigma "/\sigma '$ curves can be expected.
In fact, plotting $\sigma ''/\sigma '$ in Fig.4a, a well-defined crossing temperature
is realized. This signature for nucleation of generic superconductivity in both, the Meissner and vortex states as well was pointed out
by D. Fisher et al. \cite{FFH91} and is utilized here to trace the transition line. As demonstrated by Fig.4b, the 
crossing point persists up to the maximum available field of 19T. It is, of course, shifted downward in temperature but
the height at $T=T_c$, $\sigma''/\sigma' = 3.7(2)$, remained unchanged. Both features will be discussed in section 4.

\section{Fluctuation Dynamics and 'Generic' Transition Lines}

Since the spatial extension of the thermal fluctuations diverges at the continuous transition,
$\xi (T) =\xi (0)|1-T/T_g|^{-\nu}$, all macroscopic quantities at or near to equilibrium should be invariant
against scaling by some power of this correlation length $\xi$, like, for example, the relaxation time of the 
long-wavelength fluctuations, $\tau \sim \xi ^z$. Scaling has been proposed for the phase $\sigma "/\sigma '$  
\cite{FFH91} by using the scaled frequency, $\omega \tau$.
\begin{figure}[b]
\begin{center}
\psfig{figure=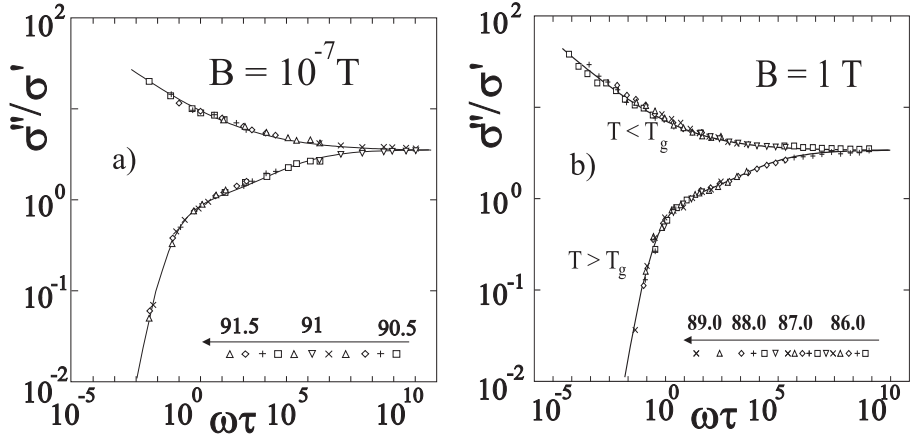,width=120mm}
\hfill
\caption{Dynamical scaling of  the phase  of the dynamic 
conductivity for the pristine film a) at zero field, where  $T_g(0)=T_c=90.85 K$ and
b) at $B=1 Tesla$, where  $T_g=86.4 K$. }
\label{FIG5}
\end{center}
\end{figure}
They collapse onto two branches of the data for above and below $T_g$ is demonstrated in Fig.5 for $\sigma (\omega)$ 
measured in zero field and in B=1T.  In addition to the scaling property we note that the presence of the high
vortex density did not change the shape of the two scaling functions, $\sigma '' /\sigma ' = P_{\pm}(\omega \tau ) $.
Moreover, the critical slowing down of the relaxation rate 
\begin{equation}
\tau^{-1}(T,B)=\tau_0^{-1} t^{\nu z} 
\label{Eq6}
\end{equation}
can be described by a single empirical form, if the reduced temperature difference $t(T,B) = |T-T_g (B)|/(T_0 - T)$ is introduced suggested in refs. \cite{KKNBA94,KNBBGB94}.
By this choice, both the exponent $\nu$z = 9.5(3) and also the amplitude $\tau _0^{-1} = 0.8\cdot  10^{12} s^{-1}$ remain
the same on either side of $T_g$ and for  {\em all}  applied fields with $T_0 = T_g (0) + 0.50K$ as the 'mean-field' temperature.

All these features provide strong evidence that the onset of generic superconductivity along $T_g (B)$ is triggered by
a common mechanism independent of the field induced vortex density.
Since the phase angle at $T_g$ is given by
\begin{equation}
\varphi_g = arctan \frac{\sigma''(T_g)}{\sigma'(T_g)} = \frac{\pi}{2} (1-z^{-1}),
\label{Eq7}
\end{equation}
this also explains the field independence of  $\varphi  _g$, i.e. z=5.7(3) in Fig.4. Moreover,
the Kronig-Kramers relations imply a power law at $T=T_g$,  $\sigma \sim (i \omega \tau_0)^{1/z-1}$
 and a  scaling behaviour of the modulus, if 
$|\sigma (\omega ,T)|$ is scaled  
by the appropriate static limits, which are here $\sigma _0 (T>T_g(B) )\sim t^{-s}$ with s=7.7 (see Fig.2b) and 
$\sigma _0 (T<T_g )\sim t^{\nu }/\omega $ with $\nu =1.7(1)$.
The unversality of the critical exponents for the vortex state has been emphasized in Ref. \cite{KNBBGB94} based on
numerous reports for thin films, which now include the 
zero-field limit for YBCO \cite{NGSWKK97} and, amazingly, also ac-data from 150nm thin indium film \cite{OK97}. A change of the universality class, however, was realized after implementation of columnar pinning 
centers by heavy-ion irradiation with a dose $n_c = 2\cdot  10^{11} cm^{-2}$
\cite{NRSWWJK96}. Here we extend this study to higher and lower matching fields, $B_\phi  = n_c \phi _0$.
\begin{figure}[b]
\parbox{70mm}{\psfig{figure=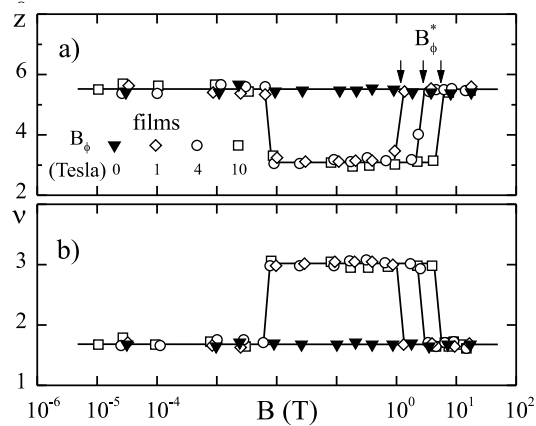,width=70mm}}\ \ \parbox{47mm}{
\begin{minipage}[t]{47mm}{\caption{Field dependence of the critical exponents $\nu$  and z determining the divergences of the correlation length and the relaxation times associated with the transition to generic superconductivity.}}
\end{minipage}}
\label{FIG6}
\end{figure}
\begin{figure}[t]
\parbox{70mm}{\psfig{figure=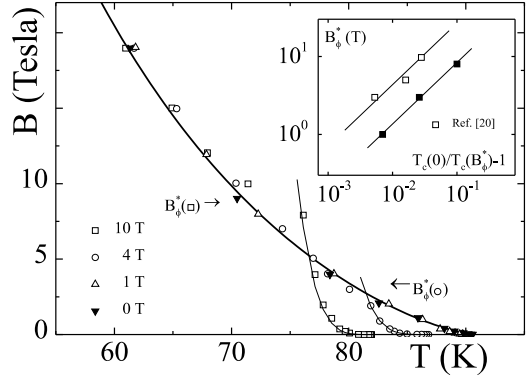,width=70mm}}\ \ \parbox{47mm}{
\begin{minipage}[t]{47mm}{\caption{Effects of magnetic field $\vec{B}||\vec{c}$ and columnar defects $ \vec{B}_{\phi}||\vec{c}$ on the transition temperatures $T_g$ to generic superconductivity. Inset: influence of $B_{\phi} ^*$  on $T_g$ in zero magnetic field. Full lines represent power laws being discussed in the text.}}
\end{minipage}}
\label{FIG7}
\end{figure}

At first, let us look at the critical exponents $\nu$ and z, evaluated from $\varphi_g$ (Eq.(\ref{Eq7})) and from the slowing down of $\tau ^{-1}$ (Eq.(\ref{Eq6})).
The results shown in Fig.6 clearly demonstrate significant changes through the defects, except for $B<10 mT$ to be considered 
elsewhere \cite{KN}. We find that the irradiation reduces z, but since the exponent of $\tau ^{-1}$, $\nu$z=9.5(3), remains unchanged, $\nu$ rises by the same amount as z decreases. Upon filling the defects with vortices, the exponents change rather sharply at some field $B_{\phi} ^* <B_{\phi}$, where they become identical to those of the pristine film.
We qualify $B_\phi ^*$ as an effective matching field being somewhat smaller than  $B_{\phi}$. Similar differences between $B_{\phi}$  and the actual defect densities have been reported by direct images of the pins \cite{WKU91}.

Another dramatic effect of the irradiation occurs in the generic transition temperatures shown in Fig.7. We note four significant features :  (i) the transition lines of the pristine film and the film with small pin density, $B_{\phi}  = 1 T$, fall on the same power law, indicated by the thick line:
\begin{equation}
B_g(T,B_{\phi} ) = B_g(B_{\phi} )  (T_g(B=0, B_{\phi} ) /T-1)^{\beta}
\label{Eq8}
\end{equation}
with $\beta$ = 1.33(3) and $B_g=50(2)T$.  
(ii) At larger doses, the transition temperatures are drastically reduced in contrast to what one would naively
expect for strong correlated pinning. This can be well described  by a fit to Eq.(\ref{Eq8}) with a greater exponent 
$\beta$=4.0(3), indicated by thin lines in Fig.7.   (iii)  If there are more vortex lines than pinning rods,
$B>B_{\phi} ^*$, the transition lines approaches that of the pristine film, which indicates that the
pinning by the rods becomes ineffective.  (iv)  As shown by the inset to Fig.7, the zero-field transition 
$T_g(B=0, B_{\phi})$
is suppressed by irradiation, which is a well-known feature but has not yet been explained. Accordingly,this reduction of $T_c$ obeys the power law, Eq.(\ref{Eq8}), with $\beta$=1.33(7) and the same
amplitude $B_g$=50T. This implies that the columnar defects exactly act like vortices in reducing the transition temperature. With slightly larger amplitude the same law holds for the transitions in irradiated YBCO-crystals  extracted from Ref.  \cite{SFK96}.

\section{Phase Fluctuations in the Presence of Disorder}

We start the discussion with two interesting observations made on the {\em pristine} film. The first is that the transition line $B_g (T)$, Eq.(\ref{Eq8}), turns out to follow rather closely the melting line predicted by Eq.(\ref{Eq3}): inserting the measured penetration depth, for which in the present range of temperatures we take $\lambda ((T)$ = 140nm/ $(1-T/T_c)^{1/3}$ \cite{NRSWWJK96,WNKSJHRK98},
we get for the stiffness $\epsilon _0 (T) = 980K (1-T/T_c)^{2/3}$. Then Eq.(\ref{Eq3}) yields $\beta = 4/3$, in perfect accord with the fitted temperature variation of the data in Fig.6 in fields up 20T. Moreover, by comparing the predicted amplitude $B_g = \phi _0 [c_L ^2 \epsilon _0 (0)/\gamma T_c]^2$ with the measured value of 50(2)T and using $\gamma =5$ for YBCO we obtain for the Lindemann number $c_L$ = 0.26(1). This value agrees surprisingly well with very recent numerical simulations:  (i) within the Bose-model for the vortex lines, $c_L = 0.25$ \cite{NB98}, and  (ii) also with $c_L=0.24$ from the anisotropic 3D-XY model for a London superconductor, $\lambda /\xi >> 1$ \cite{NS98a}. However, unlike the Bose model, the 3D-XY simulations also comprise the zero field limit where no external flux lines present. They indicate that the transitions in both zero and finite field are driven by the Onsager unbinding of thermally excited, closed vortex loops. The result $n_s \sim (1-T/T_c)^{0.67}$~ \cite{NS98a} is fully consistent with the measured penetration depth by which we explained the exponent of the generic line $\beta  = 4/3$. For $B=0$, Williams \cite{W99} arrived at essentially the same variation of $n_s$ and interpreted it by a blowout at $T_c = T_g (0)$ due to a percolation of vortex loops determined by the sample size R. At finite B, the externally induced vortex-lines appear to mediate a percolation already between small loops, thus reducing the transition temperature $T_m (B)$. 

 With this in mind it is suggestive also to attribute the reduction of the zero-field transition in the {\em columnar} defected films (see inset to Fig.7) to the thermally excited loops. We assume, that the rods act as nucleation centers for the loops which save there some nucleation energy. Then the same amplitude of $50 T$ describing the vortex-induced and rod-induced reduction of $T_c$ indicates that the saving is about the same. The suppression of $T_c (B_\phi )$ is described by the dephasing criterion, Eq.(\ref{Eq3}), if in the phase cutoff, $d_c = c_L^2 a_r /\gamma$, the mean distance between the rods $a_r$ instead of the mean vortex distance $a_0$ is used.

In finite applied fields, the transition temperature of the columnar defected films would stick at $T_c(B_\phi )$ up to $B = B_\phi ^*$ if all induced vortex lines would localize at the rods. Obviously this is not the case so that we consider the possibility of thermal depinning. Here we base our discussion on the Bose model since it was argued recently \cite{KN98}, that  for $B>0$ some results like $c_L$ turn out to be the same in the vortex-loop and  the Bose-model.  Accordingly, we use for the depinning from rods a phase cutoff at $d_r = r_0/\pi  \gamma $  \cite{BFGLV94,NV93}. By inserting this and the mean pin radius in our films, $r _0 = 3.5nm$,   into Eq.(\ref{Eq1}), one finds $T_{dp} = 0.8T_c = 72 K$, so that in fact the delocalization is occuring in the temperature range of interest, see Fig.7. However, since the BG-work treats the thermally smeared pin potential only as a perturbation of the vortex-vortex interaction for $B>B_{c1}$, the transition line is always shifted to {\em above} the melting line (see Fig.1b), while our data approach $B_g(T,B_\phi ) = B_m (T)$ from below. We believe that our results should be analyzed in a more consistent picture, which takes into account the phase breaking by thermal loops in the presence of columnar pins {\em and} externally induced vortices . One step into this direction has been made by Wallin and Girvin \cite{WG93} who obtained from simulations based on isotropic vortex loops $\nu  = 1.0(1)$ and z = 6.0(5) . Interestingly, these values are fairly close to our low-field results, but they clearly fail above 10mT, i.e. in the regime where $B_g(T)$ rises steeply with $\beta =4$ in Eq.(\ref{Eq8}). On the other hand, the agreement with the exponents of the pristine film indicates that there intrinsic linear defects nucleate the generic transition.

The most obvious consequence of the increase of the exponents over z = 3/2 \cite{W99} and $\nu = 2/3$ \cite{NS98a} for a clean 3D-XY system is the broadening of the fluctuation regime. This is illustrated for the dc resistance by Fig.2b for which one expects $\rho_0 \sim (T-T_c)^{\nu (z-1)}$. Qualitatively spoken this broadening may arise from an increased nucleation of vortex loops below $T_g$ and from a reduction of these phase fluctuations above $T_g$ by the intrinsic disorder. Similar effects on the correlation length i.e. on $\nu$ have been observed earlier on structures consisting of weakly Josephson-coupled Nb-grains \cite{LRP88}.  To best of our knowledge, the interplay between the thermally excited loops, field induced vortices, and the real disorder in high-$T_c$ materials has not yet been worked out.

To conclude, we provided evidence that thermally excited vortex loops determine the nucleation of generic superconductivity of YBCO-films in zero and finite magnetic fields. This follows from the identical scaling behaviour of the dynamic conductivity for $B=0$ and $B>0$ and the quantitative description of the generic transition line by the 3D-XY, i.e. vortex-loop model. Additional support for the dominance of these phase fluctuations comes from the observation~\cite{KN} that the relevant temperature variable, which described the critical behaviour up to vortex densities B = 19 T, is given by 
$t = | T-T_g(B) | / (T_0 - T)$ (Eq.(\ref{Eq6}). Such behaviour emerges from the temperature variation of the coupling energy between the phases in the 3D-XY Hamiltonian \cite{NS98a,LRP88}.

The author is much indebted to Dr. G. Nakielski for the fruitful collaboration during the experimental stage of the project. He also gratefully acknowledges the support by the groups of Prof. U. Merkt (Hamburg) for providing the films and of Dr. G. Wirth (GSI Darmstadt) for the irradiation, by Dr. A. G. M. Jansen (MPI Grenoble) for extending the high magnetic fields, by Dr. E. H. Brandt (MPI Stuttgart) for communicating results prior to publication, and by Dr. D. G\"{o}rlitz (Hamburg) for editing the manuscript.

\end{document}